\documentclass[sigconf]{acmart}
\newcommand{\modelname}{\textsf{DGRec}\xspace}

\AtBeginDocument{%
  \providecommand\BibTeX{{%
    \normalfont B\kern-0.5em{\scshape i\kern-0.25em b}\kern-0.8em\TeX}}}

\copyrightyear{2023}
\acmYear{2023}
\setcopyright{acmlicensed}\acmConference[WSDM '23]{Proceedings of the Sixteenth ACM International Conference on Web Search and Data Mining}{February 27-March 3, 2023}{Singapore, Singapore}
\acmBooktitle{Proceedings of the Sixteenth ACM International Conference on Web Search and Data Mining (WSDM '23), February 27-March 3, 2023, Singapore, Singapore}
\acmPrice{15.00}
\acmDOI{10.1145/3539597.3570472}
\acmISBN{978-1-4503-9407-9/23/02}
\usepackage{multirow}
\usepackage{xspace}
\usepackage{color}
\usepackage{amsmath}
\DeclareMathOperator*{\argmax}{arg\,max}

\begin{document}

\title{DGRec: Graph Neural Network for Recommendation with Diversified Embedding Generation}
\author{Liangwei Yang}
\authornote{The work is finished during internship at Applied Machine Learning (AML) of ByteDance Inc.}
\affiliation{%
  \institution{University of Illinois at Chicago}
  \city{Chicago}
  \country{USA}}
\email{lyang84@uic.edu}

\author{Shengjie Wang}
\affiliation{%
  \institution{ByteDance Inc.}
  \city{Seattle}
  \country{USA}}
\email{shengjie.wang@bytedance.com}

\author{Yunzhe Tao}
\affiliation{%
  \institution{ByteDance Inc.}
  \city{Seattle}
  \country{USA}}
\email{yunzhe.tao@bytedance.com}

\author{Jiankai Sun}
\affiliation{%
  \institution{ByteDance Inc.}
  \city{Seattle}
  \country{USA}}
\email{jiankai.sun@bytedance.com}

\author{Xiaolong~Liu, Philip S. Yu}
\affiliation{%
  \institution{University of Illinois at Chicago}
  \city{Chicago}
  \country{USA}}
\email{{xliu262,psyu}@uic.edu}

\author{Taiqing Wang}
\affiliation{%
  \institution{ByteDance Inc.}
  \city{Seattle}
  \country{USA}}
\email{taiqing.wang@bytedance.com}

\renewcommand{\shortauthors}{Liangwei et al.}

\begin{abstract}
Graph Neural Network (GNN) based recommender systems have been attracting more and more attention in recent years due to their excellent performance in accuracy. Representing user-item interactions as a bipartite graph, a GNN model generates user and item representations by aggregating embeddings of their neighbors. 
However, such an aggregation procedure often accumulates information purely based on the graph structure, overlooking the redundancy of the aggregated neighbors and resulting in poor diversity of the recommended list.
In this paper, we propose diversifying GNN-based recommender systems by directly improving the embedding generation procedure.
Particularly, we utilize the following three modules: submodular neighbor selection to find a subset of diverse neighbors to aggregate for each GNN node, layer attention to assign attention weights for each layer, and loss reweighting to focus on the learning of items belonging to long-tail categories. Blending the three modules into GNN, we present \textbf{DGRec} (\textbf{D}iversified \textbf{G}NN-based \textbf{Rec}ommender System) for diversified recommendation. Experiments on real-world datasets demonstrate that the proposed method can achieve the best diversity while keeping the accuracy comparable to state-of-the-art GNN-based recommender systems. We open source DGRec at \textcolor{blue}{\url{https://github.com/YangLiangwei/DGRec}}.
\end{abstract}




\maketitle

\section{Introduction}
We live in an era of information overflow~\cite{bigdata}, with data created every moment too large to digest in time. Recommender systems~\cite{graph_learning_RS,Wang2014VSRANK,WangAdapting2012,Sun2012LTR} target mitigating the problem by providing people with the most relevant information in the massive data. 
Recommender systems play an essential role in our daily life, such as the news feed~\cite{npa}, music suggestions~\cite{musicrec}, online advertising~\cite{dr2021}, and shopping recommendations~\cite{shoprec}.
To maximize the utility of recommendation systems, accuracy is often the only criterion  measuring how likely the users would interact with  given items. 
Companies and researchers have been building sophisticated methods~\cite{wang2022contrastvae,consisrec} to optimize accuracy during all steps in recommender systems.

However, a well-designed recommender system should be evaluated from multiple perspectives, e.g. diversity~\cite{dgcn}. Accuracy can only reflect correctness, and pure accuracy-targeted methods may lead to the echo chamber/filter bubble~\cite{echochamber} effects, trapping users in a small subset of familiar items without exploring the vast majority of others. To break the filter bubble, diversification in recommender systems is receiving increasing attention. Through an online A/B test, research~\cite{cb2cf} shows that the number of users' engagements and the average time spent greatly benefit from diversifying the recommender systems. Diversified recommendation targets increase the dissimilarity among recommended items to capture users' varied interests. 
Nevertheless, optimizing diversity alone often leads to decreases in accuracy. Accuracy and diversity dilemma~\cite{dilemma} reflects such a trade-off. Therefore, diversified recommender systems aim to increase diversity with minimal costs on accuracy~\cite{dgcn,dum,dpp}.

Graph-based recommender systems~\cite{graph_learning_RS} have attracted more and more research attention. Graph-based methods have several advantages. Representing users' historical interactions as a user-item bipartite graph can give us easy access to high-order connectivities.
Graph neural network~\cite{gnn} is a family of powerful learning methods for graph-structured data~\cite{he2021fedgraphnn,LiangSDM2018}. 
The common practice of graph-based recommender systems is designing suitable graph neural networks to aggregate information from the neighborhood of every node to generate the node embedding.
This procedure also provides opportunities for diversified recommendation~\cite{dgcn}. 
Firstly, the user/item embedding is easily affected by its neighbors, and we can manipulate the choice of neighbors to obtain a more diversified embedding representation.
Secondly, the unique high-order neighbors of each user/item node can provide us with personalized distant interests for diversification, which can be naturally captured by stacking multiple GNN layers.

\begin{figure}
    \begin{center}
    \includegraphics[width=.23\textwidth]{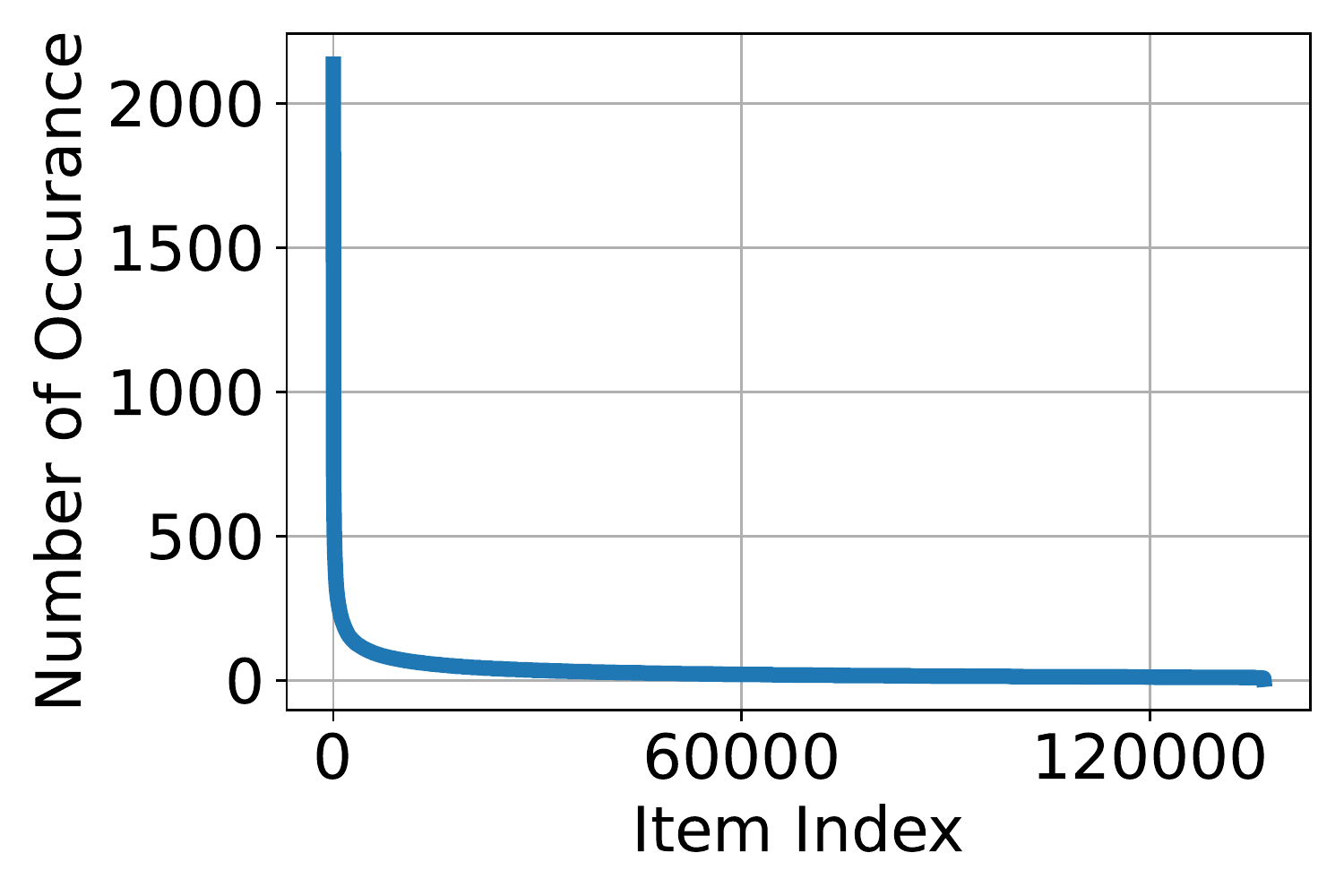}
    \includegraphics[width=.23\textwidth]{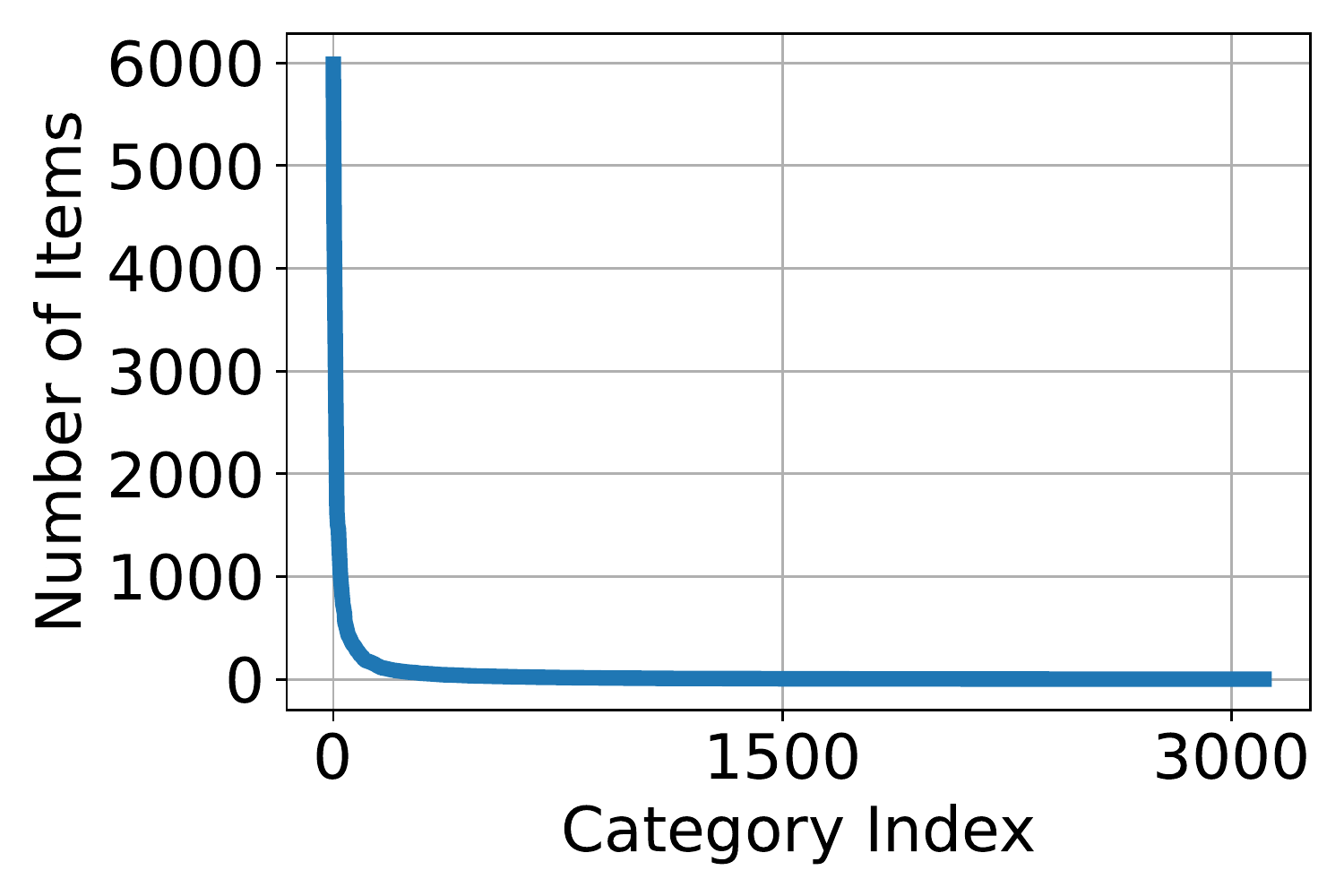}
    \end{center}
    \caption{Long tail distribution in Recommender System on TaoBao dataset~\cite{dgcn}.}
    \label{fig:log_tail}
\end{figure}

Achieving diversified recommendations using GNNs comes with the following challenges.
Firstly, how to effectively manipulate the neighborhood to increase diversity is still an open question. The popular ones will submerge the long-tail items if we have a direct aggregation on all neighbors. 
Secondly, the over-smoothing problem~\cite{oversmoothing} occurs when directly stacking multiple GNN layers. 
Over-smoothing would lead to similar representations among nodes in the graph, dramatically decreasing the accuracy performance. 
Thirdly, as seen in Figure~\ref{fig:log_tail}, the item occurrence in data and the number of items within each category both follow the power-law distribution. 
Training under such distribution would focus on the popular items/categories, which only constitute a small part of the items/categories. 
Meanwhile, the long-tail items/categories are un-perceptible during the training stage. 
Researches in graph-based diversified recommendation is very limited. Early endeavors~\cite{dilemma} assign different probabilities on edges to boost the information flow of long-tail items. DGCN~\cite{dgcn} is the first work to diversify over graph neural networks. It fails to consider the high-order connectivities and the long tail categories.

In this paper, we propose \modelname to cope with the previously mentioned challenges. We design the following three modules. \textbf{1. Submodular neighbor selection} firstly integrates submodular optimization into GNN. It
finds a diversified subset of neighbors by optimizing a submodular function. Information aggregated from the diversified subset can help us uncover the long-tail items and reflect them in the aggregated representation.
\textbf{2. Layer attention} aims to handle the over-smoothing problem. It stabilizes the training on deep GNN layers and enables \modelname to take advantage of high-order connectivities for diversification. 
\textbf{3. Loss reweighting} reduces the weight on popular items/categories. It assists the model in focusing more on the long-tail items/categories.
Our contributions are summarized as follows:

\begin{itemize}
    \item We design three modules for the diversified recommendation and propose \modelname that achieves the best trade-off between accuracy and diversity.
    \item The three modules can be easily applied to graph neural network based methods to increase recommendation diversity with a small cost on accuracy.
    \item We conduct extensive experiments on real-world datasets to show the effectiveness of \modelname and the influences of different modules.
\end{itemize}

The remaining paper is organized as follows. Section~\ref{sec:pre} gives the required preliminaries. Section~\ref{sec:method} illustrates \modelname in detail and the three proposed modules for diversification. Section~\ref{sec:experiment} conducts extensive experiments to evaluate the effectiveness of \modelname, and discusses the influence of different modules. Section~\ref{sec:related} represents the most related works for reference, and we conclude \modelname and discuss future research directions in Section~\ref{sec:conclustion}.

\section{Preliminaries}\label{sec:pre}
This section introduces some work preliminaries, including task formulation, graph neural network, and accuracy-diversity dilemma.

\subsection{Problem Statement}

For diversified recommendation task, we have a set of users $\mathcal{U} = \{u_1,u_2,...,u_{\left | \mathcal{U} \right|}\}$, a set of items $\mathcal{I} = \{i_1,i_2,...,i_{\left | \mathcal{I} \right|}\}$, and a mapping function $C(\cdot)$ that maps each item to its category. The observed user-item interactions can be represented as an interaction matrix $\textbf{R} \in \mathbb{R^{\left| \mathcal{U} \right| \times \left| \mathcal{I} \right|}}$, where $R_{u,i}=1$ if user $u$ has interacted with item $i$, or $R_{u,i}=0$ otherwise. For a graph based recommender model, the historical interactions are represented by a user-item bipartite graph $\mathcal{G}=(\mathcal{V}, \mathcal{E})$, where $\mathcal{V}= \mathcal{U}\cup \mathcal{I}$ and there is an edge $e_{u,i} \in \mathcal{E}$ between $u$ and $i$ if $R_{u,i}=1$.

Learning from the user-item bipartite graph $\mathcal{G}$, a recommender system aims to recommend top $k$ interested items $\{i_1, i_2,..., i_k\}$ for each user $u$. The diversified recommendation task requires the top $k$ recommended items to be dissimilar to each other. The dissimilarity (or diversity) of a recommended list is usually measured by the coverage of recommended categories $|\cup_{i \in \{i_1,\ldots,i_k\}} C(i)|$~\cite{coverage,dgcn}. 

\subsection{Graph Neural Network}
A Graph Neural Network is a deep learning model that operates on graph structures, and it has achieved great success in the application of many real-world tasks with graph-structured data, including social networks~\cite{group,liu2022federated}, email networks~\cite{liu2022bond} and user-item interaction graphs in recommender systems~\cite{yang2022large}.
A GNN model learns the representations of node embeddings by aggregating information from their neighbors, so that connected nodes in the graph structure tend to have similar embeddings.
The operation of a general GNN computation can be expressed as follows:
\begin{align}
    \mathbf{e}_{u}^{(l+1)} = \mathbf{e}_{u}^{(l)} \oplus \text{AGG}^{(l+1)} (\{ \mathbf{e}_{i}^{(l)} \mid i \in \mathcal{N}_u \}),
\end{align}
where $\mathbf{e}_{u}^{(l)}$ indicates node $u$'s embedding on the $l$-th layer, $\mathcal{N}_u$ is the neighbor set of node $u$, $\mbox{AGG}^{(l)}(\cdot)$ is a function that aggregates neighbors' embeddings into a single vector for layer $l$, and $\oplus$ combines $u$'s embedding with its neighbor's information. AGG$(\cdot)$ and $\oplus$ can be simple or complicated functions.

\subsection{Submodular Function}
A submodular function is a set function defined on a ground set $V$ of elements: $f: 2^V \rightarrow \mathbb{R}$. The key defining property of submodular functions is the diminishing-returns property, i.e.,
\begin{equation}
    f(v|A) \geq f(v|B) \quad\forall A \subset B \subset V, \ v \in V \text{  and  } v \notin B.   
\end{equation}
Here we use a shorthand notation $f(v|A) := f(\{v\} \cup A) - f(A)$ to represent the gain of an element $v$ conditioned on the set $A$. The diminishing-returns property naturally describes the diversity of a set of elements, and submodular functions have been applied to various diversity-related machine learning tasks with great success in practice, such as text summarization, sensor placement, and training data selection~\cite{zheng2014submodular,hoi2006batch}. Submodular functions are also applied as a re-ranking method to diversify recommendations, which is orthogonal to the relevance prediction model.
Submodular functions also exhibit nice theoretical properties to be solved with strong approximation guarantees using efficient algorithms~\cite{nemhauser1978analysis}.

\section{Method}\label{sec:method}
\begin{figure*}
      \begin{center}
        \includegraphics[width=1\textwidth]{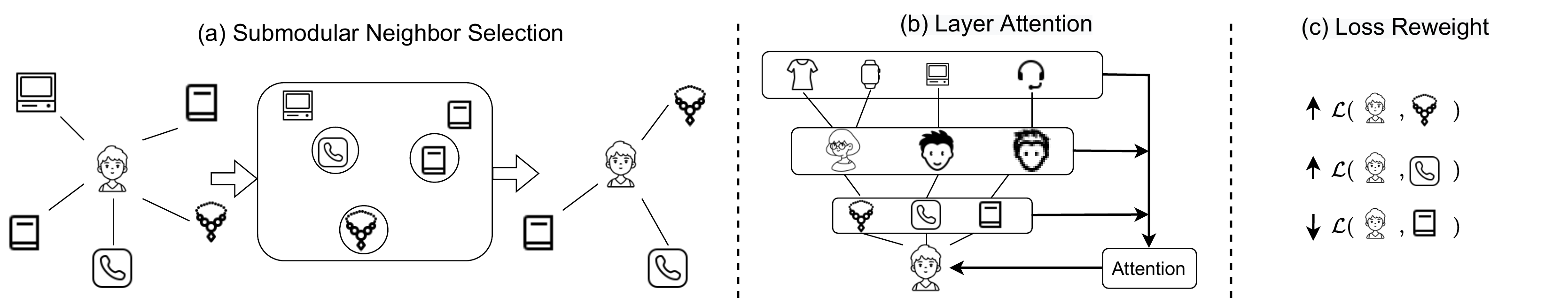}
      \end{center}
        \caption{The framework of \modelname. (a) Submodular neighbor selection module (Section~\ref{sec:neighbor selection}) finds a diversified subset of neighbors on the embedding space for aggregation. (b) Layer attention module (Section~\ref{sec:layer attention}) alleviates the over-smoothing problem from high-order connections. (c) Loss reweighting module (Section~\ref{sec:loss reweight}) adjusts weight for each sample to focus on the training of long-tail categories.}
        \label{framework}
\end{figure*}

In this section, we first present the backbone GNN-based recommender system of \modelname, and then illustrate the three modules to obtain diversified recommendations during the embedding generation procedures. The framework of \modelname is shown in Figure~\ref{framework}. More specifically, it consists of the following components: Submodular neighbor selection, Layer attention and Loss reweighting.

\subsection{Overall Training Framework}
Based on the user-item bipartite graph $\mathcal{G}$, a GNN-based recommender system generates user/item embeddings by graph neural networks to predict user's preference.

\subsubsection{Embedding Layer}
Similar to the learning representation of words and phrases, the embedding technique is also widely used in recommender systems~\cite{lightgcn,mfbpr}: an embedding layer is a look-up table that maps the user/item ID to a dense vector:
\begin{equation}
    \textbf{E}^{(0)}=\left(\textbf{e}_1^{(0)}, \textbf{e}_2^{(0)}, \ldots, \textbf{e}_{|\mathcal{U}|+|\mathcal{I}|}^{(0)}\right), \\
\end{equation}
where $\mathbf{e}^{(0)}\in \mathbb{R}^d$ is the $d$-dimensional dense vector for user/item. An embedding indexed from the embedding table is then fed into a GNN for information aggregation. Thus it is noted as the "zero"-th layer output $\mathbf{e}_i^{(0)}$.

\subsubsection{Light Graph Convolution}
We utilize the light graph convolution~\cite{lightgcn} (LGC) as the backbone GNN layer. It abandons the feature transformation and nonlinear activation, and directly aggregates neighbors' embeddings, and is defined as:
\begin{equation}
\begin{aligned}
      \mathbf{e}_u^{(l+1)}=\sum_{i\in \mathcal{N}_u}\frac{1}{\sqrt{|\mathcal{N}_u|}\sqrt{|\mathcal{N}_i|}}\mathbf{e}_i^{(l)},\\  
    \mathbf{e}_i^{(l+1)}=\sum_{u\in \mathcal{N}_i}\frac{1}{\sqrt{|\mathcal{N}_i|}\sqrt{|\mathcal{N}_u|}}\mathbf{e}_u^{(l)},
\end{aligned}
\end{equation}
where $\mathbf{e}_u^{(l)}$ and $\mathbf{e}_i^{(l)}$ are user $u$'s and item $i$'s embedding at the $l$-th layer, respectively. $\frac{1}{\sqrt{|\mathcal{N}_u|}\sqrt{|\mathcal{N}_i|}}$ is the normalization term following GCN~\cite{gcn}. $\mathcal{N}_u$ is $u$'s neighborhood that selected by submodular function as illustrated in Section~\ref{sec:neighbor selection}. Each LGC layer would generate one embedding vector for each user/item node. Embedding generated from different layers are from the different receptive field. The final user/item representation is obtained by layer attention illustrated in Section~\ref{sec:layer attention}:
\begin{equation}
\begin{aligned}
    \mathbf{e}_u=\text{Layer\_Attention}\left(\mathbf{e}_u^{(0)},\mathbf{e}_u^{(1)},\ldots,\mathbf{e}_u^{\text{(layer num)}}\right), \\
    \mathbf{e}_i=\text{Layer\_Attention}\left(\mathbf{e}_i^{(0)},\mathbf{e}_i^{(1)},\ldots,\mathbf{e}_i^{\text{(layer num)}}\right). \\
\end{aligned}
\end{equation}

\subsubsection{Model Optimization}
After we obtain $\mathbf{e}_u$ and $\mathbf{e}_i$, the score of $u$ and $i$ pair is calculated by dot product of the two vectors. For each positive pair $(u,i)$, a negative item $j$ is randomly sampled to compute the Bayesian personalized ranking (BPR)~\cite{mfbpr} loss. To increase recommendation diversity, we propose to reweight the loss to focus more on the long-tail categories:
\begin{equation}\label{loss}
\mathcal{L}=\sum_{(u,i)\in \mathcal{E}}w_{C(i)}\mathcal{L}_{bpr}(u,i,j) + \lambda {\left \| \Theta \right \|}_2^2,
\end{equation}
where $w_{C(i)}$ is the weight for each sample based on its category, which is illustrated in Section~\ref{sec:loss reweight}. $\lambda$ is the regularization factor. $j$ is a randomly sampled negative item.

\subsection{Submodular Neighbor Selection}\label{sec:neighbor selection}
In GNN-based recommender systems, user/item embedding is obtained by aggregating information from all neighbors. 
Popular items would overwhelm the long-tail items. In Figure~\ref{framework}(a), the user's embedding would be much more similar to books if we aggregate all the neighbors. 
At the same time, the necklace information is overwhelmed in the user's representation. 
The submodular neighbor selection module aims to select a set of diverse neighbors for aggregation.
In our setting of GNN neighbor selection, the ground set for a user node $u$ consists of all of its neighbors $\mathcal{N}_u$.
Facility location function~\cite{facility_location} is a widely used submodular function that evaluates the diversity of a subset of items by first identifying the most similar item in the selected subset $\mathcal{S}_u$ to every item $i$ in the ground set ($\max_{i' \in \mathcal{S}_u}\text{sim}(i,i')$ $\forall i \in \mathcal{N}_u \backslash \mathcal{S}_u$) and then summing over the similarity values. 
Intuitively, a subset with a high function value indicates that for every item in the ground set, there exists a similar item in the selected subset, or in other words, the selected subset is very diverse and representative of the ground set.
The facility location function is formally defined as follows:
\begin{equation}\label{facility}
    f(\mathcal{S}_u) = \sum_{i\in \mathcal{N}_u\backslash\mathcal{S}_u}\max_{i' \in \mathcal{S}_u}\text{sim}(i,i'),
\end{equation}
where $\mathcal{S}_u$ is the selected neighbor subset of user $u$, and $\text{sim}(i,i')$ is the similarity of $i$ and $i'$, which is measured by Gaussian kernel parameterized by a kernel width $\sigma^2$:
\begin{equation}
    \text{sim}(i,i')=\text{exp}\left(-\frac{||\mathbf{e}_i - \mathbf{e}_{i'}||^2}{\sigma^2}\right).
\end{equation}
$\mathcal{S}_u$ is constrained to having no greater than $k$ items for some constant $k$, i.e., $|\mathcal{S}_u| \leq k$. Maximizing the submodular function (\ref{facility}) under cardinality constraint is NP-hard, but it can be approximately solved with $1-e^{-1}$ bound by the greedy algorithm~\cite{nemhauser1978analysis}. The greedy algorithm starts with an empty set $\mathcal{S}_u := \emptyset$, and adds one item $i \in \mathcal{I}\backslash\mathcal{S}_u $ with the largest marginal gain to $\mathcal{S}_u$ every step:
\begin{equation}
\begin{aligned}
    &\mathcal{S}_u \leftarrow \mathcal{S}_u \cup i^*, \\
    & i^* = \argmax_{i \in \mathcal{N}_u\backslash\mathcal{S}_u}\ [f(\mathcal{S}_u \cup i) - f(\mathcal{S}_u)].
\end{aligned}
\end{equation}
After $k$ steps of greedy neighbor selection, we can obtain the diversified neighborhood subset of each user. The subset is then used for aggregation. 
We also note that our framework works for any choice of a submodular function. We choose the facility location function as it is generally applicable to numerical features (with certain similarity metric). We also discuss other choices of submodular functions in the empirical studies.

\subsection{Layer Attention}\label{sec:layer attention}
Different GNN layers generate embeddings based on information from different subsets of nodes:
the $l$-th layer would aggregate from the $l$-th hop neighbors. 
We can reach a diversified embedding by aggregating from the high-order neighbors. However, the direct stack of several GNN layers would cause the over-smoothing problem~\cite{oversmoothing}. As shown in Figure~\ref{framework}(b), layer attention is designed in \modelname to increase diversity by high-order neighbors and mitigate the over-smoothing problem at the same time.

For each user/item, we have $L$ embeddings generated by $L$ GNN layers. Layer attention aims to get the final representation by learning a Readout function on $[\mathbf{e}^{(0)},\mathbf{e}^{(1)},\ldots,\mathbf{e}^{(L)}]$ by attention~\cite{liu2021deep}:
\begin{equation}
     \mathbf{e} = \text{Readout} ([\mathbf{e}^{(0)},\mathbf{e}^{(1)},\ldots,\mathbf{e}^{(L)}]) 
    = \sum_{l=0}^{L} a^{(l)}\mathbf{e}^{(l)},
\end{equation}
where $a^{(l)}$ is the attention weight for $l$-th layer. It is calculated as:
\begin{equation}
    a^{(l)} = \frac{\text{exp}(\langle\mathbf{W}_{\text{Att}}, \mathbf{e}^{(l)}\rangle)}{\sum_{l'=0}^{\text{L}} \text{exp}(\langle\mathbf{W}_{\text{Att}}, \mathbf{e}^{(l')}\rangle)}.
\end{equation}
Here $\mathbf{W}_{\text{Att}}\in \mathbb{R}^{d}$ is the parameter for attention computation. The attention mechanism can learn different weights for GNN layers to optimize the loss function. It can effectively alleviate the over-smoothing problem~\cite{liu2021deep}.


\subsection{Loss Reweighting}\label{sec:loss reweight}
As shown in Figure~\ref{fig:log_tail}, the number of items within each category is highly imbalanced and follows the power-law distribution. A small number of categories contains the most items while leaving the large majority of categories with only a limited number of items. Training the model by directly optimizing the mean loss over all samples would leave the training of long-tail categories imperceptible. In \modelname, we propose to reweight the sample loss during training based on its category. As shown in Figure~\ref{framework}(c), \modelname would decrease the weight relatively if the item belongs to popular categories, and increase the weight relatively if it belongs to long-tail categories. 

In practice, we borrow the idea of class-balanced loss~\cite{classbalancedloss} to reweight the sample $(u,i)$ based on the category effective number of items. The weights $w_{C(i)}$ in Equation~\ref{loss} are calculated by:
\begin{equation}
    w_{C(i)} = \frac{1-\beta}{1-\beta^{|C(i)|}},
\end{equation}
where $\beta$ is the hyper-parameter that decides the weight. A larger $\beta$ would further decrease the weight of popular categories.

\section{Experiment}\label{sec:experiment}
In this section, We conduct extensive experiments on two real-world datasets to answer the following research questions~(RQs):
\begin{itemize}
    \item \textbf{RQ1}: Does \modelname outperform existing methods in the diversified recommendation?
    \item \textbf{RQ2}: How do the hyper-parameters influence \modelname, and how can we trade off accuracy and diversity in \modelname?
    \item \textbf{RQ3}: Are the three components in \modelname necessary to boost diversification?
    \item \textbf{RQ4}: What is the influence of different submodular functions?
\end{itemize}

\subsection{Experimental Setup}
\subsubsection{Datasets}
To evaluate the effectiveness of \modelname, we conduct experiments on two real-world datasets with category information. The statistics of the two datasets are shown in Table~\ref{data stat}.
\begin{itemize}
    \item \textbf{TaoBao}~\cite{dgcn}: This dataset contains users' behavior on TaoBao platform, which was provided by Alimama\footnote{https://github.com/tsinghua-fib-lab/DGCN/tree/main/data}. This dataset contains users' multiple kinds of behaviors, including clicking, purchasing, adding items to carts, and item favoring. All those behaviors are treated as positive samples. To ensure the quality of the dataset, the 10-core setting is adopted, i.e., only users/items with at least $10$ interactions are retained.
    \item \textbf{Beauty}~\cite{liang2021enhancing}: This dataset contains product review information and metadata from Amazon\footnote{http://jmcauley.ucsd.edu/data/amazon/links.html}. Following the setting in~\cite{liang2021enhancing}, the 5-core version is adopted to ensure data quality.
\end{itemize}

For both datasets, we randomly split out $60\%$ for training, $20\%$ for validation, and $20\%$ for testing. Validation sets are used for hyper-parameter tuning and early stopping. We report results on the test set as the final results.

\begin{table}
  \caption{Statistics of the Datasets}
  \label{data stat}
  \begin{tabular}{c c c}
        \toprule
        Dataset & TaoBao & Beauty \\
        \hline
        Users & 82,633 & 8,159 \\
        Items & 136,710 & 5,862 \\
        Interactions & 4,230,631 & 98,566 \\
        \hline
        Categories & 3,108 & 41 \\
        Average Category Size & 43.986 & 139.595\\
        \bottomrule
  \end{tabular}
\end{table}

\subsubsection{Baselines}
To empirically evaluate and study \modelname, we compare our model with representative recommender system baselines. Note that \modelname is compatible with the re-ranking-based methods such as DPP~\cite{dpp}, MMR~\cite{mmr}, DUM~\cite{dum} and Diversified PMF~\cite{dpmf}. Thus we do not compare those methods in the experiments. Selected baselines are shown as follows:

\begin{itemize}
    \item \textbf{Popularity}: It is a non-personalized recommendation method that only recommends popular items to users.
    \item \textbf{MF-BPR}~\cite{mfbpr}: It factorizes the interaction matrix into user and item latent factors.
    \item \textbf{GCN}~\cite{gcn}: It is one of the most widely used graph neural networks.
    \item \textbf{LightGCN}~\cite{lightgcn}: It is the state-of-the-art recommender system. LightGCN is a GCN-based model but removes the transformation matrix, non-linear activation, and self-loop.
    \item \textbf{DGCN}~\cite{dgcn}: It is the current state-of-the-art diversified recommender system based on GNN.
\end{itemize}

\subsubsection{Evaluation Metrics} Following previous works~\cite{dgcn,dpp,Cheng2017WWW}, we use two different kinds of metrics to evaluate the accuracy and diversity respectively. We aim to get a diversified item set during the retrieval stage, so Recall and Hit Ratio (HR) are used to measure the accuracy. Coverage is used to measure diversity, which counts the number of covered categories of recommended items. To save space, we only report Top-100 and Top-300 retrieval results. We can reach the same conclusion for other top-N retrievals.

\subsubsection{Parameter Setting}
In experiments, we tune all the baselines using the validation set and report the results on the test set. Adam~\cite{adam} is used as the optimizer.
Following the setting of DGCN, we fix the embedding size to be $32$ and randomly sample $4$ negative items for each positive user-item pair for a fair comparison. 
Other hyper-parameters are tuned by grid search.
Early stopping is utilized to alleviate the over-fitting problem. We stop training if the performance on validation set does not improve in $10$ epochs. \looseness=-1

\subsection{Performance Evaluation (RQ1)}

\begin{table*}[htbp]
\caption{Overall comparison on TaoBao dataset, the best and second-best results are in bold and underlined, respectively.}
\label{tab:comparison_taobao}
\begin{tabular}{lcccccc}
\toprule

\multirow{2}{*}{Method} & \multicolumn{6}{c}{TaoBao} \\

\cmidrule(r){2-7}
& Recall@100 & Recall@300 & HR@100 & HR@300 & Coverage@100 & Coverage@300 \\
\hline
Popularity & 0.0186 & 0.0357 & 0.1496 & 0.2562 & \underline{38.2449} & 75.9837 \\
MF-BPR~\cite{mfbpr} & \underline{0.0487} & \underline{0.0971} & \underline{0.3103}  & \underline{0.4889} & 34.0812 & 71.8802\\
GCN~\cite{gcn} & 0.0446  & 0.0923 & 0.2840  & 0.4634  & 37.2577 & 79.2985\\
LightGCN~\cite{lightgcn} & \textbf{0.0528} & \textbf{0.1063} & \textbf{0.3261} & \textbf{0.5097}  & 32.7069 & 69.3502  \\ \hline
DGCN~\cite{dgcn} & 0.0394 & 0.0831 & 0.2634 & 0.4369 & 38.1183 & \underline{84.4989} \\
\textbf{\modelname}  & 0.0472  & 0.0951 & 0.3026 & 0.4817 & \textbf{39.0597} & \textbf{89.1684} \\

\bottomrule 
\end{tabular} 
\end{table*}

\begin{table*}[htbp]
\caption{Overall comparison on Beauty dataset, the best and second-best results are in bold and underlined, respectively.}
\label{tab:comparison_beauty}
\begin{tabular}{lcccccc}
\toprule

\multirow{2}{*}{Method} & \multicolumn{6}{c}{Beauty} \\

\cmidrule(r){2-7}
& Recall@100 & Recall@300 & HR@100 & HR@300 & Coverage@100 & Coverage@300 \\
\hline
Popularity & 0.1012 & 0.2096  & 0.1833 & 0.3124 & 16.0213  & \textbf{27.9336} \\
MF-BPR~\cite{mfbpr} & 0.2310 & 0.3863 & 0.3404 & 0.4966 & 15.8728 & 25.6659 \\
GCN~\cite{gcn} & 0.2388 & 0.3897 & \underline{0.3423} & 0.3897 & 16.5311 & 25.5634  \\
LightGCN~\cite{lightgcn} & \textbf{0.2517} & \textbf{0.4205} & \textbf{0.3688} & \textbf{0.5318} & 15.0203 & 23.9421 \\ \hline
DGCN~\cite{dgcn} & 0.2395 & 0.3790 & 0.3418 & 0.4792 & \underline{18.2876} & 26.9694 \\
\textbf{\modelname}  & \underline{0.2399} & \underline{0.3915} & 0.3420 & \underline{0.5021} & \textbf{19.0557} & \underline{27.5704} \\

\bottomrule 
\end{tabular} 
\end{table*}

\begin{figure}
      \begin{center}
        \includegraphics[width=.45\textwidth]{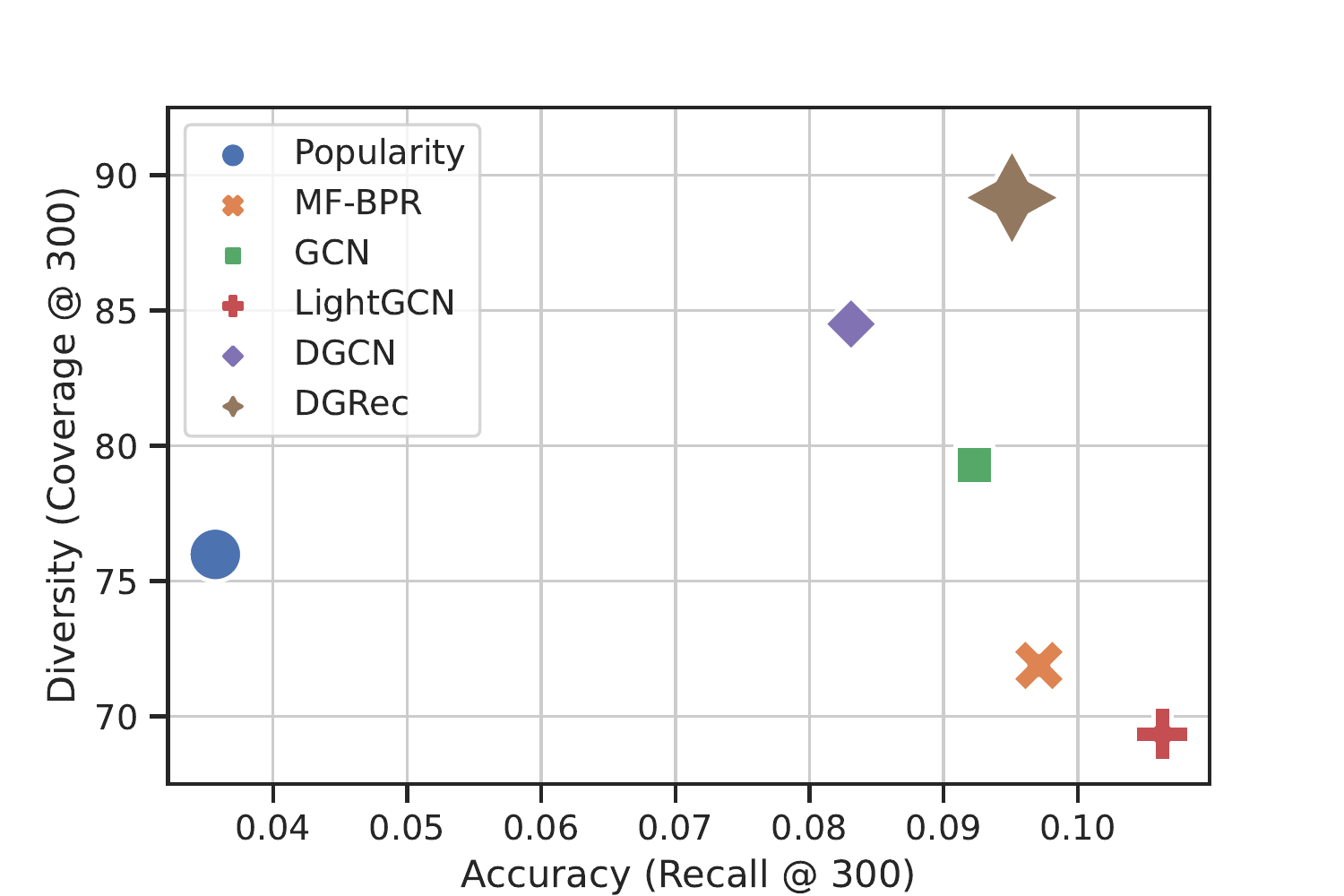}
      \end{center}
        \caption{Accuracy-Diversity trade-off comparison on TaoBao dataset. The upper-right model is enlarged.}
        \label{trade-off_comparison}
\end{figure}

We report the experiment results in Table~\ref{tab:comparison_taobao} (TaoBao dataset) and Table~\ref{tab:comparison_beauty} (Beauty dataset). We have the following observations:
\begin{itemize}
    \item \modelname generally achieves the best on Coverage@100 and Coverage@300 except being second to Popularity in terms of Coverage@300 on the Beauty dataset. Considering Coverage@300 on the Beauty dataset, \modelname is just slightly lower than Popularity. It shows that \modelname can achieve the most diversified recommendation results. 
    \item Though LightGCN always achieves the best Recall and Hit Ratio, its Coverage is always the lowest. It shows that LightGCN can not achieve an accuracy-diversity balance.
    \item While achieving the best Coverage, \modelname has similar results with the second best on Recall and Hit Ratio. It shows \modelname increases the diversity with a small cost on the accuracy, which well balances the accuracy-diversity trade-off. 
    \item \modelname surpasses DGCN on all metrics. It shows \modelname surpasses the SoTA model, and the design of \modelname is superior in both accuracy and diversity.
\end{itemize}

To make a clearer comparison of all methods, we illustrate the accuracy-diversity trade-off in Figure~\ref{trade-off_comparison}. Accuracy and diversity are measured by Recall@300 and Coverage@300, respectively. We can clearly observe that \modelname stands in the most upper-right position, which shows \modelname achieves the best trade-off. Compared with \modelname, all other models with similar accuracy (GCN, MF-BPR) have an obvious drop in diversity. Compared with LightGCN, \modelname greatly increases diversity with a small sacrifice on accuracy.

\subsection{Parameter Sensitivity (RQ2)}
In this section, we study the influence of different hyper-parameters on \modelname, and how to trade-off between accuracy/diversity.
\subsubsection{Layer Number.}

\begin{figure}
    \begin{center}
    \includegraphics[width=.235\textwidth]{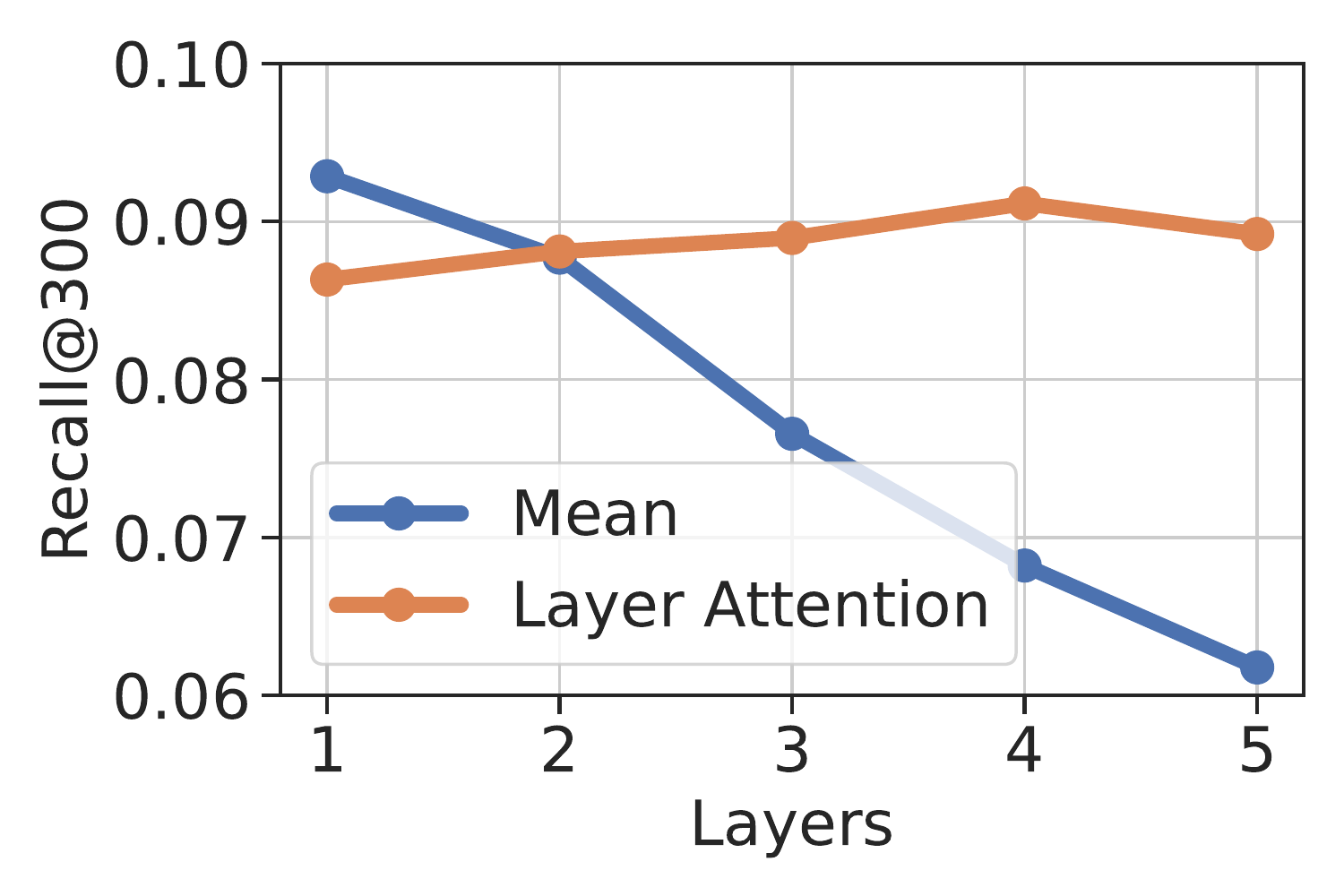}
    \includegraphics[width=.235\textwidth]{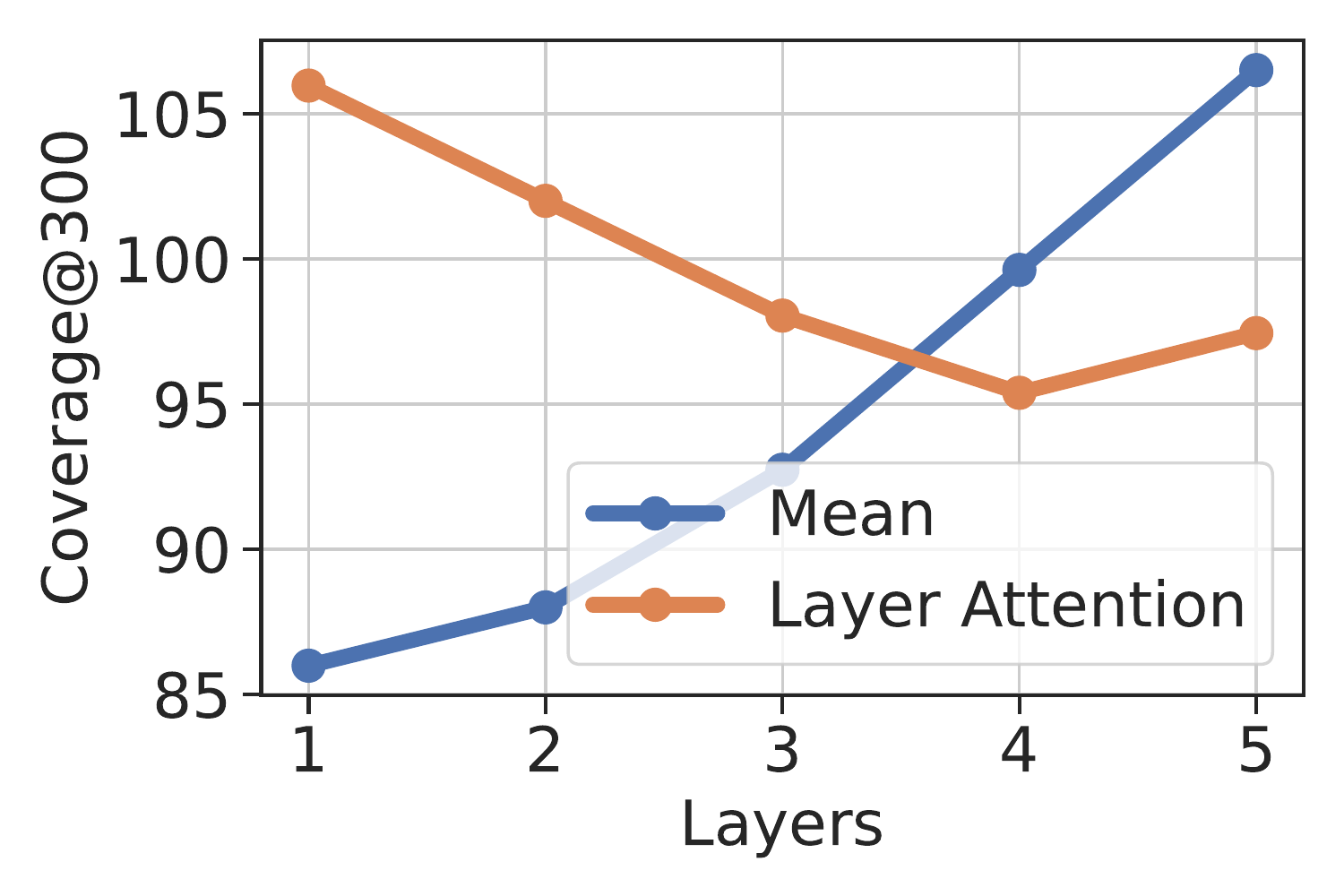}
    \end{center}
    \caption{Layer combination experiments on TaoBao dataset. Mean combines embedding learned from different layers by mean average. Layer attention combines these embeddings by the attention module illustrated in Section~\ref{sec:layer attention}.}
    \label{fig:layers}
\end{figure}

The layer number is an influential hyper-parameter in the GNN-based recommender system,
which indicates the number of GNN layers stacked to generate the user/item embedding. 
We compare our proposed layer attention with the mean aggregation~\cite{lightgcn} on both accuracy and diversity. Experimental results are shown in Figure~\ref{fig:layers}. 
With the mean aggregation, we can see Recall@300 drops quickly with the increase of layers. It reflects the well-known over-smoothing problem~\cite{oversmoothing} in GNN. 
The increase in Coverage@300 verifies our hypothesis that we can obtain a diverse embedding representation by adding more information from higher-order connections. 
However, mean aggregation does not make an effective trade-off between accuracy and diversity.
The sharp drop on Recall@300 makes the increased diversity meaningless. With the proposed layer attention, \modelname does not suffer from the over-smoothing problem and achieves gradually increased Recall@300 with the increase of layers. It shows layer attention can effectively learn different attention weights for each layer to fit the data.
At the same time, \modelname generally achieves a high Coverage@300. It shows the layer attention module can retain a good performance on diversity with a different number of layers.
When mean aggregation and layer attention achieve similar Recall@300 (2 layers), Coverage@300 of layer attention is much larger than mean aggregation. The case is similar if we compare Recall@300 when they achieve similar Coverage@300.
It shows layer attention used in \modelname can achieve a much better accuracy diversity trade-off than mean aggregation.

\subsubsection{Hyper-parameter $\beta$.}

\begin{figure}
      \begin{center}
        \includegraphics[width=.45\textwidth]{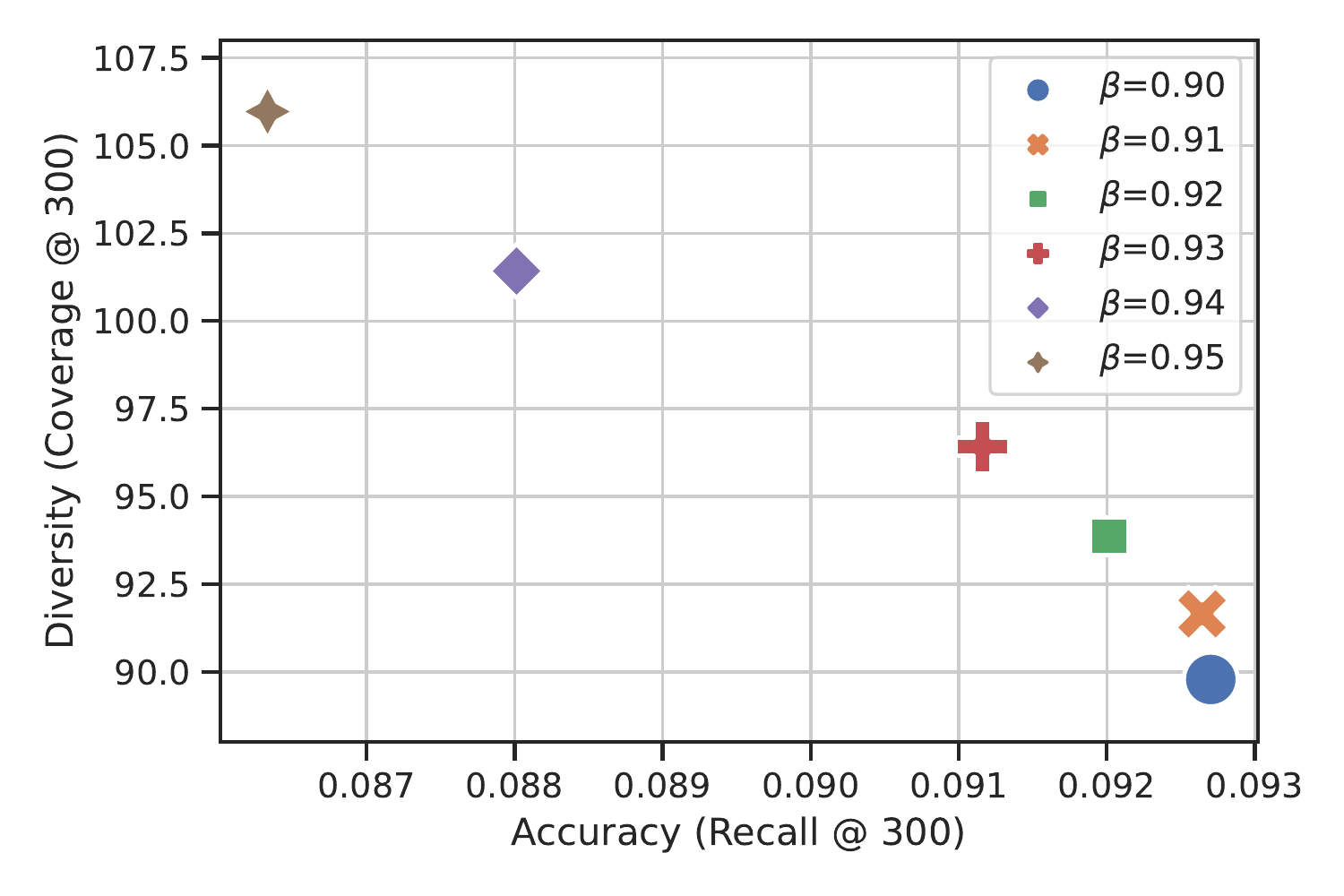}
      \end{center}
        \caption{Accuracy-diversity trade-off by loss reweighting.}
        \label{fig:beta}
\end{figure}

This hyper-parameter is introduced in Section~\ref{sec:loss reweight} to control the weight on loss calculated on each sample. With a larger $\beta$, \modelname would concentrate more on the items that belong to long-tail categories. The accuracy-diversity trade-off diagram is shown in Figure~\ref{fig:beta}. With the increase of $\beta$, accuracy gradually drops, and diversity increases. It indicates focusing on the training of long-tail categories can greatly increase diversity. We can also observe that the accuracy drops slowly with the increase in diversity. When $\beta=0.95$, \modelname achieves a Coverage@300 of more than $105$ and Recall@300 of more than $0.086$. Experimental results show that by focusing on the training of items belonging to the long-tail categories, $\beta$ can be used effectively to balance between diversity and accuracy.

\subsubsection{Hyper-parameter $\sigma$ and $k$}
\begin{figure}
    \begin{center}
    \includegraphics[width=.23\textwidth]{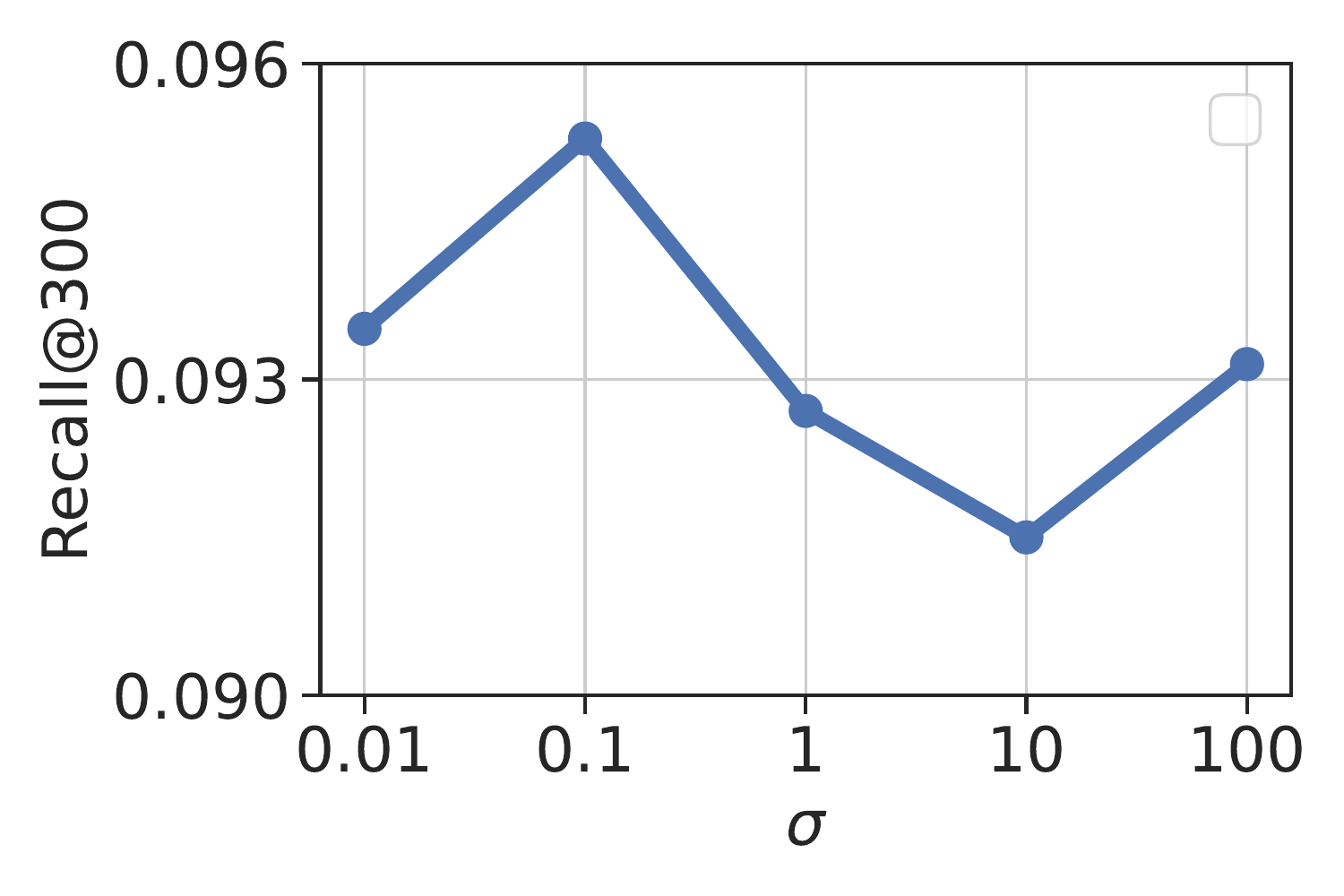}
    \includegraphics[width=.23\textwidth]{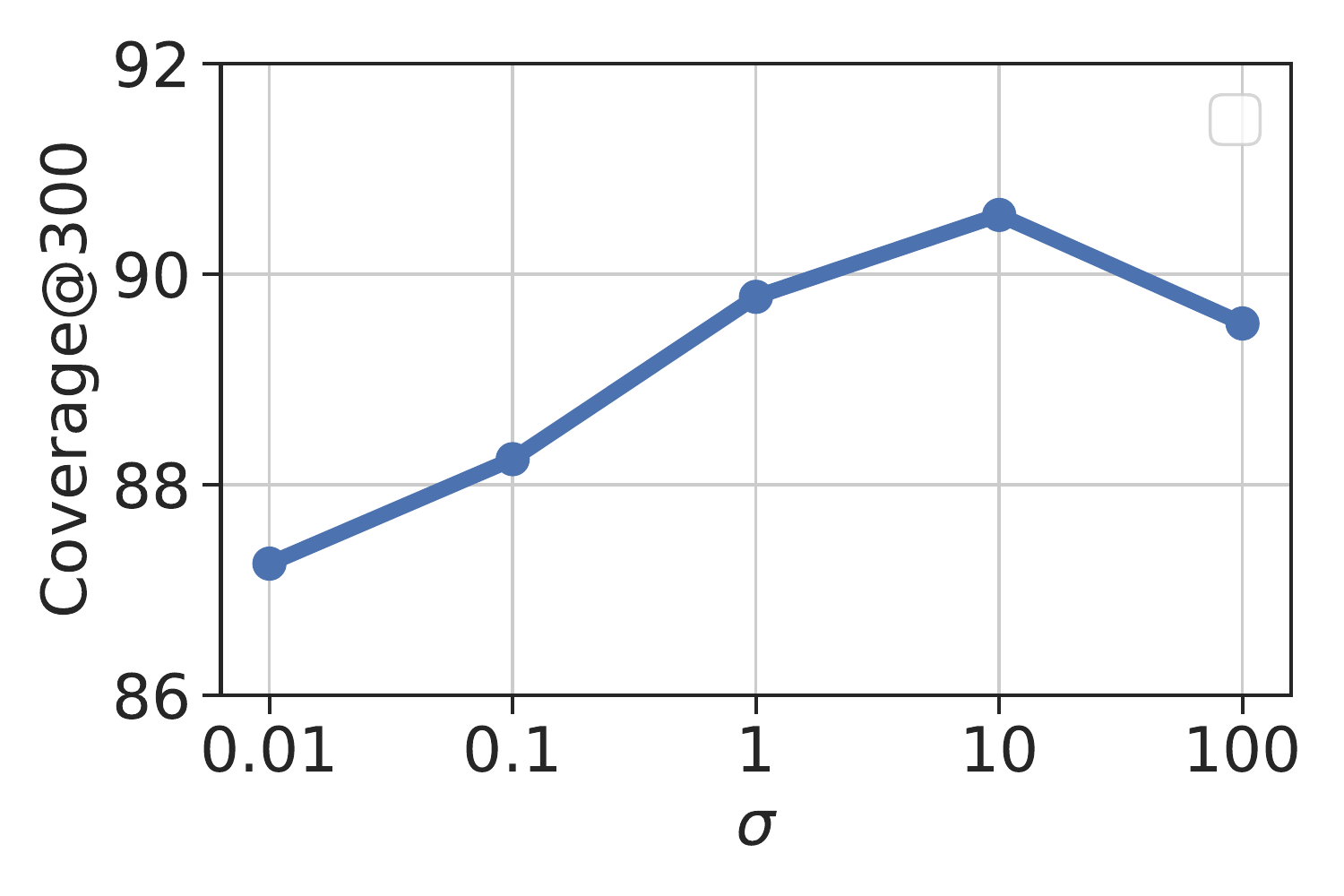}
    \includegraphics[width=.23\textwidth]{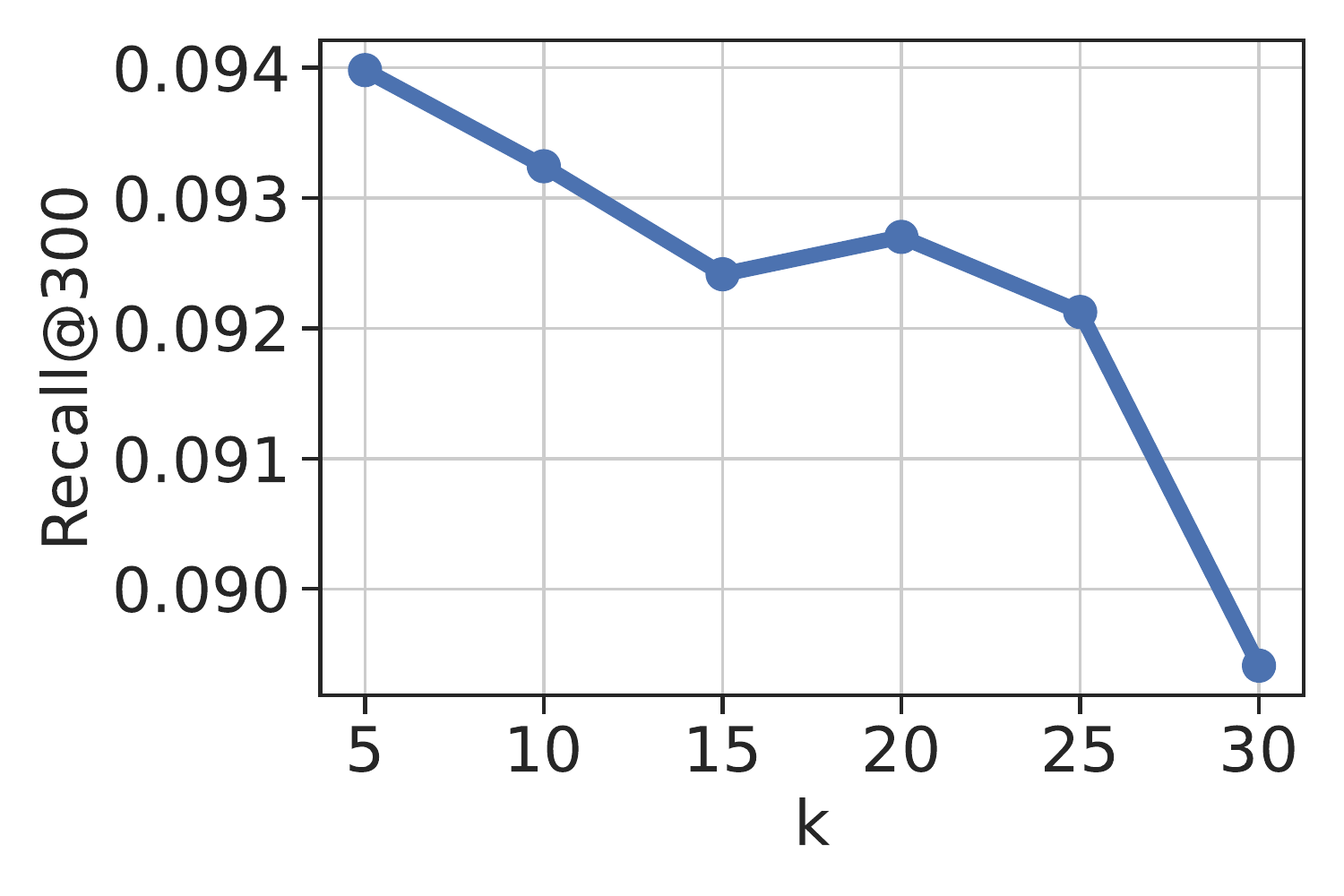}
    \includegraphics[width=.23\textwidth]{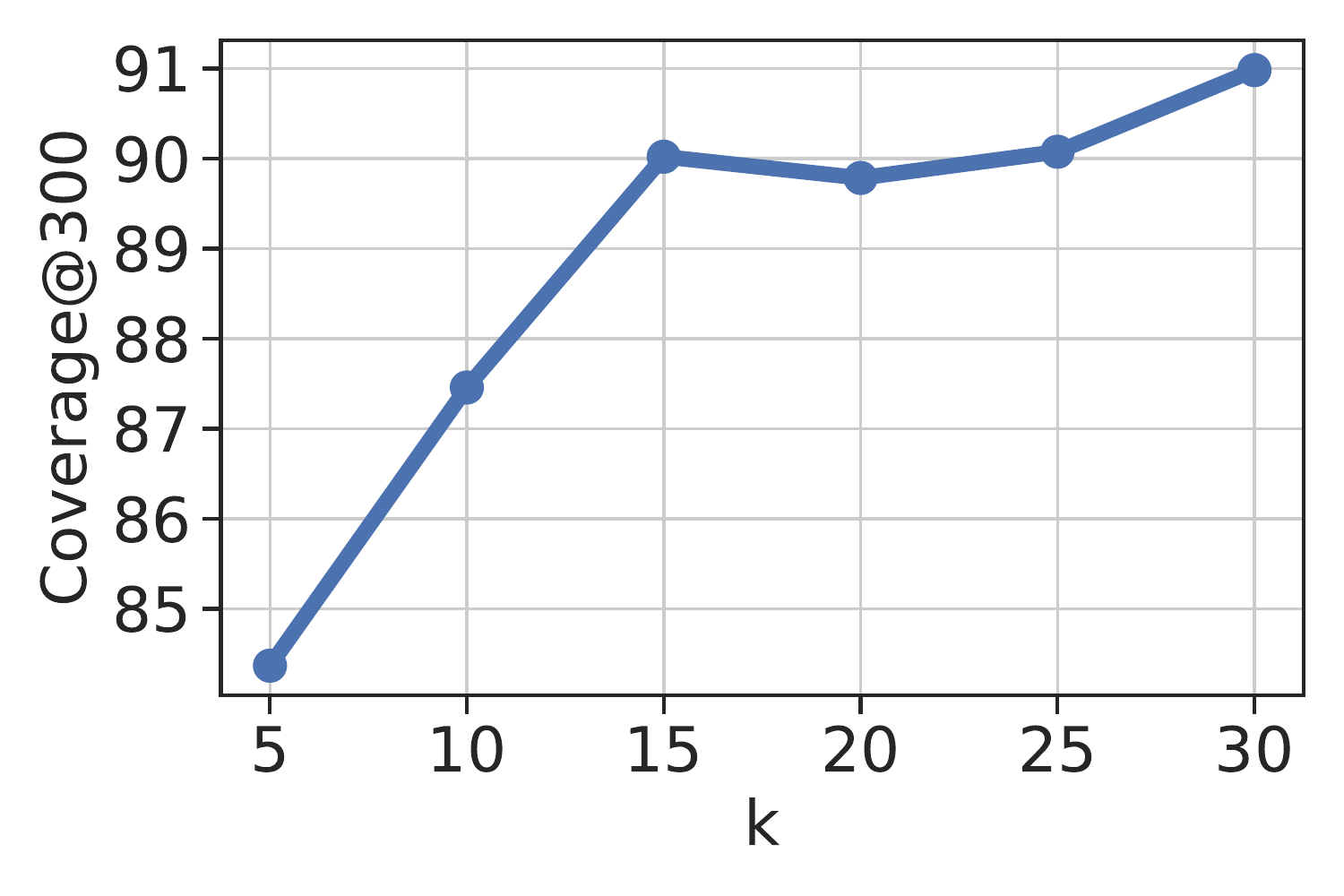}
    \end{center}
    \caption{Parameter sensitivity of $\sigma$ and $k$. $\sigma$ controls the similarity computation and $k$ is the number of selected neighbors.}
    \label{fig:sensitivity}
\end{figure}

$\sigma$ and $k$ are introduced in Section~\ref{sec:neighbor selection}. $k$ is the budget for neighbor selection, and $\sigma$ is used to compute the pair-wise similarity of neighbors. Experimental results are shown in Figure~\ref{fig:sensitivity}. 

We can observe that \modelname is not that sensitive to $\sigma$. \modelname has a stable good performance on both Recall@300 and Coverage@300 with $\sigma$ varies from $0.01$ to $100$. With different $\sigma$, we can also see the trade-off between accuracy/diversity. When Coverage@300 achieves the best at $10$, Recall@300 is the worst.

$k$ is the number of neighbors for GNN aggregation. Neighbors are selected by submodular function to maximize diversity. As we can see from Figure~\ref{fig:sensitivity}, Recall@300 gradually decreases, and Coverage@300 increases with the increase of $k$. Submodular neighbor selection selects a diversified subset of neighbors. With a larger set, \modelname can aggregate from more diversified neighbors, which would lead to an increase in diversity. At the same time, accuracy would drop as a trade-off. We can also observe that Recall@300 does not drop much with the increase in diversity.

Experiments on $\sigma$ and $k$ show \modelname is not sensitive to the submodular selection module, and \modelname would not have a dramatic change because of this module. Meanwhile, this module can also balance accuracy and diversity by $\sigma$ and $k$.

\subsection{Ablation Study (RQ3)}
In this section, we perform an ablation study on the TaoBao dataset by removing each of the three modules. Experiment results are shown in Table~\ref{tab:ablation}. We can have the following observations:
\begin{itemize}
    \item The intact \modelname achieves the best C@300. The combination of proposed modules can effectively increase diversity.
    \item When we remove the submodular neighbor selection module, C@300 drops from $89.1684$ to $84.9129$ while there is only a tiny difference on Recall@300 and HR@300. It shows the submodular neighbor selection module can increase the diversity with minimal cost on accuracy.
    \item When we remove the layer attention module, C@300 decreases with the increase on R@300 and HR@300. It indicates layer attention balances accuracy and diversity.
    \item When we remove the loss reweighting module, R@300, HR@300, and C@300 all drop greatly. The loss reweighting module has the largest impact on \modelname because it not only balances the training on long-tail categories but also guides the learning of layer attention.

\end{itemize}

\begin{table}
  \caption{Ablation study on TaoBao dataset. We show~\modelname' performance when removing each of the modules.}
  \label{tab:ablation}
  \begin{tabular}{l c c c}
        \toprule
        Method & R@300 & HR@300 & C@300 \\
        \hline
        \modelname & 0.0951   & 0.4817   & 89.1684  \\
        $\quad$ w/o Submodular selection & 0.0982  & 0.4869 & 84.9129 \\
        $\quad$ w/o Layer attention & 0.1009  & 0.4976  & 82.9553 \\
        $\quad$ w/o Loss reweighting & 0.0886  & 0.4612   & 79.3286  \\
        \bottomrule
  \end{tabular}
\end{table}

\subsection{Choice of Submodular Functions (RQ4)}

\begin{figure}
    \begin{center}
    \includegraphics[width=.23\textwidth]{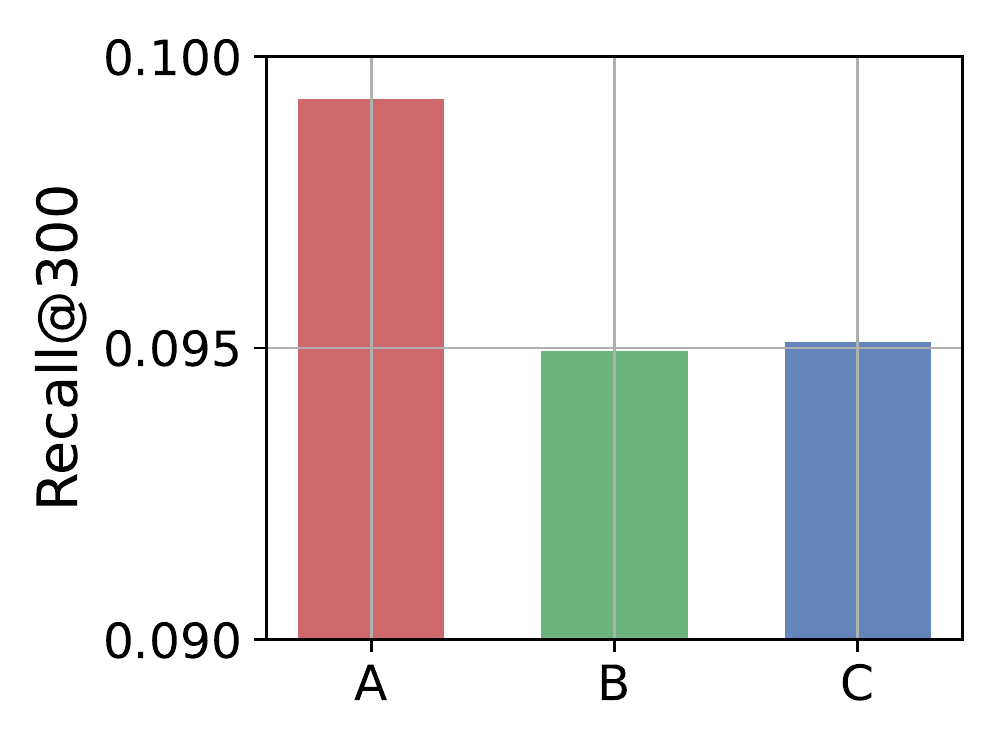}
    \includegraphics[width=.23\textwidth]{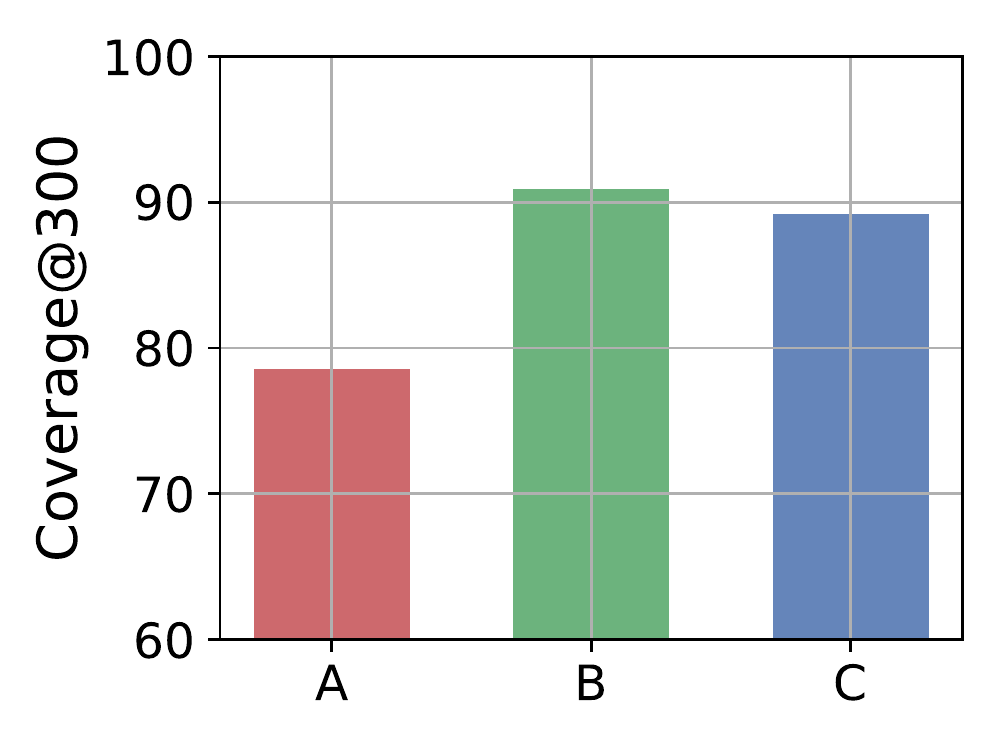}
    \end{center}
    \caption{The influence of different submodular functions. A: bucket coverage function, B: category coverage function, and C: facility location function.}
    \label{fig:subfunc}
\end{figure}

In this section, we compare the influence of different submodular functions on model performance. 
We use two commonly used submodular functions to replace the facility location function. Experimental results are shown in Figure~\ref{fig:subfunc}.
Model A utilizes bucket coverage submodular function~\cite{wei2015submodularity}. Before selection, it clusters on each dimension and divides each dimension into buckets. The submodular function counts the gain on covered buckets.
Model B utilizes category coverage submodular function~\cite{wei2015submodularity}. This function counts the gain on covered categories.
Model C is \modelname, which utilizes the facility location function. Among the three models, model A and model C do not need item category information. They directly select neighbors based on neighbor embedding. Model B needs item category information to be able to compute category coverage gain during each selection.

From Figure~\ref{fig:subfunc}, we observe that compared with the other two models, model A has much higher performance on Recall@300 and much lower performance on Coverage@300. It shows the selection of submodular functions has an influential impact on performance. Model B and model C achieve similar results with respect to Recall@300 and Coverage@300. It indicates the embedding learned by model C can accurately capture the category information, and the facility location function enlarges the category coverage during neighbor selection. We select the facility location function in \modelname for two reasons. Firstly, it can nearly achieve the best diversity compared with other methods. Secondly, it does not need category information during aggregation, which can enlarge the application scenarios when the category information is unobserved.

\section{Related Work}\label{sec:related}
In this section, we introduce the related work of DGRec, which includes Graph Neural Network based recommender system and diversified recommendation.

\subsection{Graph Neural Network based Recommender System}
With GNN showing excellent performance on graph-structured data, GNN-based recommender systems~\cite{graph_learning_RS} are attracting more and more attention. These methods represent the user's historical interactions as a user-item bipartite graph with easy access to high-order connectivity. 
GCMC~\cite{gcmc} utilizes encoder-decoder structure on graph to complete interaction matrix. 
 SpectralCF~\cite{spectralcf} is the first to study the spectral domain of the user-item bipartite graph. It proposes spectral convolution operation to find the latent interactions, and greatly increase the recommendation performance on cold-start items. PinSAGE~\cite{pinsage} designs a special random walk to accelerate the learning on the large-scale bipartite graph, which is applied on the Pinterest platform. NGCF~\cite{ngcf} directly aggregates information from neighbors in the bipartite graph, and explicitly injects the collaborative signal in the learned embedding. LightGCN~\cite{lightgcn} simplifies NGCF by removing the overhead computation of linear transformation and non-linear activation. The simplified operation not only achieves better performance but also reduces the training time. UltraGCN~\cite{ultragcn} takes a further step in simplifying graph convolutional network. It skips the finite layers of aggregation, and directly computes the infinite convolution stage as a constraint during training. MetaKRec~\cite{yang2022metakrec} reconstructs the knowledge graph as edges between items before graph convolution.
 
Previous GNN-based recommender systems nearly all focus on increasing accuracy while leading to poor diversity. \modelname is also built upon Graph Neural Network. The proposed three modules can be added to previous GNN-based recommender systems and make up for their diversity shortcomings.

\subsection{Diversified Recommendation}
Diversified recommendation aims to recommend users with a diversified subset of items to help users find unexplored interests. 
Diversified recommendation is first proposed by \citet{diversify1}. They use a greedy method to select items during the retrieval procedure.
\citet{dilemma} points out the accuracy/diversity dilemma, and propose HeatS/ProbS methods to choose the information propagation probability for each edge in the user/item bipartite graph.
~\citet{Cheng2017WWW} introduced a new pairwise accuracy metric and a normalized topic coverage diversity metric to measure the performance of accuracy and diversity.
Then several re-ranking-based methods are proposed to diversify recommendation lists after the retrieval procedure. DUM~\cite{dum} uses the submodular function to greedy guide the selection of item selection in the re-ranking procedure to maximize the item's utility. Diversified PMF~\cite{dpmf} computes $l_2$ loss between items as diversity. Determinantal point process (DPP)~\cite{dpp} re-ranks items to achieve the largest determinant on the item's similarity matrix. \citet{antikacioglu2017post} formulate a recommender system as a subgraph selection problem from diversified super graphs, and they use minimum-cost network flow methods to achieve a fast algorithm in diversification. \citet{teo2016adaptive} assign global/local diversification weights in the training of recommender systems. CB2CF~\cite{cb2cf} designs sliding spectrum decomposition to capture user's diversity perception over long item lists. Through online testing, CB2CF shows diversification can increase the number of engagements and time spent on the Xiaohongshu platform. DDGraph~\cite{ddgraph} selects implicit edges by quantile progressive candidate selection and re-constructs the user-item bipartite graph to increase diversity. DGCN~\cite{dgcn} is the first GNN-based diversified recommendation method. It selects node neighbors based on the inverse category frequency for diverse aggregation and further utilizes category-boosted negative sampling and adversarial learning to diverse items in the embedding space.

\modelname focuses on diversifying the GNN-based recommender system in the retrieval stage. Among the previous methods, the re-ranking-based methods such as DPP and DUM are compatible with our method. DGCN is the most similar work with \modelname. We both focus on how to increase diversity on GNN-based methods.

\section{Conclusions}\label{sec:conclustion}
In this paper, we target diversifying GNN-based recommender systems with diversified embedding generation. We design three modules that can be easily applied to GNN-based recommender systems to achieve diversification with minimal cost on accuracy.
Based on the three modules, we propose \modelname. When considering diversity, it surpasses the state-of-the-art diversified recommender system. It also achieves comparable accuracy with the most advanced accuracy-based recommender system.
\modelname enables the trade-off between accuracy and diversity by several hyper-parameters. Extensive experiments on real-world datasets illustrate the influence of different modules.



\begin{acks}
This work is supported in part by NSF under grants III-1763325, III-1909323,  III-2106758, and SaTC-1930941.
\end{acks}

\bibliographystyle{ACM-Reference-Format}
\bibliography{sample-base}

\appendix

\end{document}